\documentstyle [preprint,aps] {revtex}
\begin {document}
\draft
\title {Search for Structural Modulations in $UPt_3$ Using Laue X-ray
Diffraction}
\author {Brett Ellman, Mark Sutton, Benoit Lussier, Ralf Br\"uning\cite{ralf}
and Louis Taillefer}

\address {Department of Physics,
McGill University, 3600 University Street,
Montr\'eal, Qu\'ebec, Canada H3A 2T8}

\author {Stephen Hayden}

\address {H.H. Wills Laboratory, Physics Department, Bristol University,
Bristol,
United Kingdom}

\author {Grace Shea-McCarthy}

\address {Brookhaven National Laboratory, Bldg. 901A, DAS, Upton, NY 11973}

\maketitle

\begin {abstract}
Using synchrotron radiation-based Laue diffraction methods, we have searched
for
structural anomalies in a single-crystal whisker of
the heavy fermion superconductor $UPt_3$.  Our data cover a large
volume of k-space with high sensitivity.  We find no signs of the structural
modulations seen in transmission electron microscopy studies of whiskers
grown under identical conditions.  Our result has consequences for theories
of symmetry-broken superconductivity in this material.
\end {abstract}


Over the last several years, a ``standard model''
of sorts has arisen to explain
the multiple superconducting phases observed in $UPt_3$\cite{reviews}.
It runs thusly:  the
hexagonal crystal structure together with the anisotropic
Cooper pair coupling mechanism (the nature of which is still unclear) lead to a
two-fold degenerate superconducting
ground state.  The two zero-field transitions observed in a variety of
experiments occur when a symmetry-breaking
field lowers the lattice symmetry, lifting the degeneracy
and resulting in two distinct free
energies as a function of temperature, which cross at the second transition.
In the ``standard model'', the symmetry-breaking field is taken to
be the weak ($\approx 0.03 \mu_B$)
in-plane antiferromagnetic moment, which lowers the symmetry from
hexagonal to orthorhombic.
This scenario received significant support from pressure-temperature
phase diagram studies\cite{hpt} which showed that both the magnetic order and
the
splitting of $Tc$ are suppressed by the same hydrostatic pressure.
There are, however, important unresolved issues as shown by, e.g.,
recent neutron scattering experiments\cite{neut} which show that the
magnetic order does not behave as expected in an external field.
In 1994, Midgley {\it et al.}\cite{migdley} proposed an alternative
symmetry-breaking
field\cite{mineev}, namely, a {\it structural}
modulation observed by them in transmission electron microscope (TEM) images
of $UPt_3$.  The modulation was found to possess a complex reciprocal-space
structure which
appears to depend on preparation methods and temperature\cite{second_mig}.
TEM studies by Ellman {\it et al.}\cite{our_TEM} on single-crystal whiskers
also found a
modulation which,
at least in the a-c plane, has a particularly simple morphology.  In
particular, the k-space modulation was found to be at $\vec{q} \approx 0.1a^* +
0.1 c^*$,
where $a^*$ and $c^*$ are the hexagonal lattice vectors. (The geometry of the
whiskers, which are needles with the long axis along $\hat c$, made it hard to
search for the modulation in the other basal plane
direction.)
The modulation was observed to be spatially coherent over
microns distances and to generate sharp electron
diffraction satellite spots about the primary hexagonal Bragg peaks.
Note that this modulation would lower the crystal symmetry in
much the same fashion as the antiferromagnetic order.
If this
structural modulation is intrinsic to $UPt_3$, there are profound implications
for the model of superconductivity outlined above.  Furthermore, the nature of
the
modulation itself (e.g., a charge density wave) remains unclear.  Before a
great deal of effort is
expended in exploring the nature and implications of the structural features
observed in
TEM, however,
it is essential to verify that they do not result from TEM preparation
procedures (e.g., ion milling) and, furthermore, that they reflect a {\it
bulk} property of $UPt_3$ and not a surface feature (TEM only probes
to a depth of
order 50 nm).  The obvious technique of choice is x-ray diffraction, which, in
the transmission geometry used here, probes the entire thickness of the
sample, which is of order 30 microns.  Two
problems then arise:  (1) modulation reflections (e.g., CDW satellites)
can be quite weak and (2) TEM has poor q (energy) resolution, leading to
uncertainties in the actual reciprocal-space position of the observed
modulation.  Put simply, it is hard to know where to look when using a high
resolution triple crystal (TCS) spectrometer, and searching
the required volume of k-space
requires a great deal of time.
Synchrotron-based Laue diffraction circumvents both problems:  the intense
white radiation available from a bending magnet makes looking for weak peaks
possible
while Laue diffraction perforce samples a large volume of reciprocal space.
We note that, until recently, Laue diffraction has not often been used by
physicists,
except for the relatively undemanding task of aligning crystals.
Thus the purpose of this paper is twofold:  to
search for modulations in $UPt_3$ and to explore the utility of the venerable
Laue techniques in the age of synchrotron sources.

$UPt_3$, in the absence of modulations, possesses
the hexagonal $P6_3/mmc$ structure
with two formula units per unit cell.  The lattice constants, a = 5.764 \AA \
 and c = 4.899 \AA, are accurately known from a variety of x-ray and neutron
experiments.  The sample used in the present study was an
unannealed single crystal whisker grown in ultra-high vacuum
in an identical manner
to the crystal used in our TEM study\cite{our_TEM}.
The only difference between the two
samples is that no ion milling was performed on the x-ray crystal.  The whisker
is approximately $2 \times 0.05 \times 0.05 \, mm^3$ in size, with the
hexagonal c
axis oriented along the long (growth) direction.  Transmission
Laue experiments were
performed on NSLS beamline X26C using unfocused white radiation, a 200 $\mu$m
collimator, and an image plate detector (Fuji), while TCS
data on the same sample were taken on beamline X20A and B.  The intensity scale
for the image plate data is roughly linear and will be measured in arbitrary
"image units" (IU).

The TCS experiment
carefully searched several regions of k-space (near the (2 0 3) and (0 0 2)
Bragg reflections) for modulation peaks with very
high sensitivity, as explained below.  No evidence of a modulation was found.
However, only a comparatively small volume of k-space was searched, leaving
open the possibility that a modulation observable with TEM was missed in the
TCS search.
Anomalous structure {\it was} found in a TCS study of a Czochralski grown
sample
in the form of a broad peak near the (200) Bragg peak.  However, Laue data on
the same sample showed numerous low angle grain boundaries and other
imperfections that make interpretation of the TCS data difficult.
The whisker Laue
data clearly show the extremely high
crystallographic quality of the sample.  In Fig. 1a we
show a longitudinal TCS scan of the (200) Bragg peak.  The full width at
half maximum (FWHM) is 0.007 $c^*$,
which is 1/14 that obtained for a large Czochralski-grown sample of the type
that exhibit the standard multi-phase diagram and is comparable to the
resolution of our diffractometer.  Also
shown in Fig. 1 is a theta scan with a FWHM of 0.011 degrees, slightly more
than twice that obtained for a high-quality Si crystal.  The width of
scans along $\hat q$ measures the spread in lattice constant due to,
e.g., internal stresses, while scans
along $\theta$ probe the angular distribution from low angle
grain boundaries.  Therefore, taken together,
these scans imply that the whisker has a much smaller spread in lattice
constant than bulk samples and an (angular) mosaic of the same order as
a high-quality Si sample:  crystallographically, this is an excellent sample.

The schematic diagram in Fig. 1c shows a two-dimensional $\hat a - \hat c$-
plane slice
of the k-space geometry of the Laue experiment.  The small and large circles
represent, respectively, the minimum and maximum energies for which we see
reflections.
Whether a Bragg reflection (represented to scale by the array of dots)
is seen or not in the actual measurement depends both on the reflection's
structure factor and the beam
intensity at the appropriate energy.  In our results, we see peaks due to
photons
of about 3.5 keV to 70 keV.
Note that a Laue experiment is only sensitive to the
scattering angle $\theta$,
and thus the intensity of a reflection is the integral of the scattering along
{\bf q}, i.e.,  all higher order reflections (e.g., (200),(400),
etc.) fall onto the same spot as the fundamental peak.

The main experimental result of this paper is the 5 s exposure Laue image
shown in Fig. 2.
A diffuse background (due to scattering off the collimator, air, etc.) has been
subtracted using a two-dimensional polynomial fit on six overlapping regions
of the figure.  The dark area in the middle covers the beam stop.  Diffuse
scattering along the beam axis makes it difficult to subtract the background
near the beam stop, decreasing our sensitivity in this region.
The zones are clearly visible as ellipses including the origin.
It is essential
in searching for anomalous reflections that we know the orientation of the
sample to high accuracy.  There are six parameters:  the three Euler angles of
the crystal axes relative to the lab frame, the distance from the image plate
to the sample, and the x and y positions of the image plate center relative to
the
beam axis.  (Including additional two parameters to account for horizontal
and vertical tilts of the plate does not improve the quality of the fits.)
These constants were simultaneously determined by choosing a small set (about
20) bright
reflections whose indices could be guessed {\it a priori} and then fitting the
calculated image positions as a function of the six parameters to the
positions measured off the Laue image.  The resulting crystal orientation
angles were thereby determined to $\approx \pm$ 0.5 degree.
The agreement between the predicted and observed positions of the peaks is not
uniform across the image.
Because these deviations
are systematic and small compared to typical inter-reflection distances,
they do not impede our search for satellite peaks due to a modulation.

Since the orientation of the crystal was highly constrained by the fitting
procedure, searching for peaks additional to those accounted for by the
$P6_3/mmc$
bulk structure of $UPt_3$ is simple:  we simply plot the positions of Bragg
peaks predicted for the hexagonal lattice and ascertain whether any
measured Bragg spots are unaccounted for.  In Fig. 3 we show a selected region
of the data of Fig. 2 along with the positions of predicted reflections.
All of these data
are consistent with the calculation.
The same agreement with the calculation is found across
the entire image plane.
Thus, {\it there are no unexplained reflections}:
we are left with no sign of a modulation.
To make this statement more quantitative, we show in Fig. 4 (left panel) a
contour plot of the same region around the $(\overline{4} \; \overline{6} \;
0)$ Bragg peak as is shown in Fig. 2b.  Each peak is labeled with its
(fundamental) index.  No
peaks except for those allowed under $P6_3/mmc$ symmetry are evident.  The
lowest contour level shown (32 IU) corresponds, in terms of integrated
intensity, to approximately 1/1000 of the integrated intensity of the
$(\overline{4} \; \overline{6} \; 0)$ peak.  The noise (fluctuations in the
subtracted background) is about a factor of 2 lower than the lowest contour.
We will discuss other quantitative bounds on our sensitivity later in the text.
We have also looked in other
scattering directions (albeit with less sensitivity) with the same result.

It is important to ascertain what a modulation peak might look like if it {\it
were} present.  In Fig. 2b and 2c we show the positions of modulation peaks at
all basal-plane equivalent modulation wavevectors
$<0.1 \; \overline{0.1} \; \overline{0.1}>$
i.e., symmetry equivalent to
the modulation extracted from the whisker TEM data.  While all harmonics of a
given Bragg
peak fall onto the same point on the image, the position of the modulation
satellite relative to the Bragg peak is
dependent on the relative magnitudes of the modulation wave vector and the
Bragg peak wave vector.  For example, in Fig. 2c the observed Bragg peak is
actually the $(\overline{6}\;\overline{14}\;\overline{8})$ (the second
harmonic)
while the modulation positions have been calculated with respect to the
$(\overline{3}\;\overline{7}\;\overline{4})$ fundamental.  However, one can see
that the modulation would be visible even if its intensity were
governed by that of the dominant Bragg peak (in which case the distorted
hexagon would be a factor of 2 smaller) except perhaps for the modulation
wavevectors that result in satellites closest to the main peak.  This brings
up another important point, namely that
the position of the
satellite peaks about the main peak is
a strong function of the position of the Bragg peak on the image plane.  Since
the Bragg
peak images are not symmetric (e.g., they display diffuse streaking due
to, e.g.,
phonons),
it is essential to study peaks in various zones when searching for weak
satellites.  As an example of how a satellite would appear we show
in the right panel of Fig. 4 the data of Fig. 2b with
a modulation peak inserted by hand.
This ``fake'' is an actual peak lifted from the image, normalized to
an integrated intensity of 500 IU and then placed at image plane
coordinates corresponding to
($\overline{4} \; \overline{6} \; 0$) + ($0.1 \; \overline{0.1} \;
\overline{0.1}$).
The extra
peak is obviously not accounted for by the Laue calculation and would present
a clear signal of a structural modulation if it were present in the data.
Similarly, adding a $\vec{\delta q}$ = (0 $x$ 0) (this is the direction
unconstrained by the TEM measurements) with arbitrary $x$ to the
modulation k-vector
moves the satellite further from the main peak and makes it even more obvious.

We have also looked in other
scattering directions (albeit with less sensitivity) with the same result.
For example, in Fig. 5 we show the Laue image obtained for a transverse
scattering
geometry, i.e., the image plate at 90 degrees with respect to the
incident beam.  Analysis of the data using exactly the same techniques as the
longitudinal case showed that all of the peaks were well accounted for by the
$P6_3/mmc$ bulk structure of $UPt_3$.

One cannot set a single quantitative limit on the minimum satellite intensity
we could observe.  It is more difficult to detect a weak peak near an
intense reflection than near another weak one since the satellite is
superimposed on the tail of the Bragg peak.  Furthermore, a modulation may
or may
not result in satellites whose intensities scale with those of the parent
peaks.
Therefore we
will simply give one measure of sensitivity.  The integrated intensity of
the brightest peaks (measured using a 20 ms exposure to minimize
saturation of the image plate)
is approximately
$5 \times 10^7$ IU in 5 seconds.  Thus the 500 count ``fake''
in Fig. 4 corresponds conservatively to a ratio
(satellite intensity)/(intensity of brightest peak) of order $1 \times
10^{-5}$.
The aforementioned diffractometer measurements placed a limit
approximately
two
orders of magnitude lower on the existence of satellites in much
smaller
regions of reciprocal space.
However, as mentioned earlier in a more general context, Laue techniques
are very sensitive to diffuse scattering insofar as a Laue peak measures
the integral of scattering along a line in k-space.  Therefore Laue
experiments are potentially more sensitive to diffuse peaks than
high resolution TCS
measurements that integrate over tiny volumes of k-space.
Of course, if the satellite peaks were {\it much} broader than the observed
Bragg
peaks,
they might (for a given integrated intensity) fall below the background.
Given the
micron-size correlation lengths seen in the whisker TEM data\cite{our_TEM},
however, we consider this unlikely.

There are a number of possible reasons the modulation is clearly observed in
TEM but not in the present experiment.  In particular, it is possible that
bulk samples are intrinsically different than the whisker examined here.
Two facts speak against this, however.
Whiskers seem to have the same superconducting phase diagram as bulk samples
insofar as
numerous whiskers grown in an
identical fashion
to our sample show features in the upper critical field that are diagnostic
of a multi-component superconducting phase diagram\cite{whiskers_ok}, as do
the bulk samples.  Furthermore, a nominally identical whisker examined in a
TEM showed a highly coherent modulation\cite{our_TEM}.  Thus we believe that
whiskers and bulk-grown material are fundamentally similar (though of higher
crystallographic quality).
Alternatively, it is possible that the modulation is introduced during the
TEM preparation or imaging processes (e.g., during ion milling or electron
bombardment).  Finally, it is possible
that our sensitivity to scattering caused by the modulation is simply not
high enough compared to a TEM.
The most important limiting factors in this measurement were the
relatively intense diffuse background and the inability to expose for longer
times
due to saturation of the plate reader.  Neither of these is a fundamental
problem:
the first may be remedied by working in helium gas along with better shielding
and collimation, while the second may
be avoided by masking the brightest Bragg reflections at the image plate.
Thus we may safely predict an eventual increase of about 2 orders of
magnitude
in our sensitivity to modulations in future experiments,
allowing the Laue measurements to equal or exceed  conventional diffractometer
sensitivities.

The detailed analysis of Laue data is no
more difficult than that of TCS data.  The broad k-space
acceptance and sensitivity to diffuse scattering make
Laue diffraction a very appealing technique
for a wide variety of experiments.  There is also another, more practical
advantage: Laue experiments allow one to probe scattering in various
k-space directions without moving the sample.  Thus studies of, e.g.,
temperature dependencies of scattering may be done without mounting an
entire cryostat or furnace on a three-circle diffractometer.

To conclude, we observe no sign that $UPt_3$ deviates from $P6_3/mmc$ symmetry.
Our study covered a large area of k-space with relatively high
sensitivity.
Our results speak against the possibility that structural modulations are
responsible for the symmetry-breaking in $UPt_3$.

\begin{figure}
\caption{(a) Longitudinal scan of the (002) Bragg peak of the 
whisker.  The FWHM resolution of the diffractometer
is comparable to the width of the peak.
(b)  Rocking curve for the same sample.  Solid lines are guides to the
eye.  Taken together, these data
indicate the high crystallographic quality of the sample.  (c)  Schematic
diagram of the Laue scattering geometry in the $\hat a$-$\hat c$ plane.  Each
dot between the smallest and
largest circle represents a reflection that may
appear in the data (see text).}
\end{figure}

\begin{figure}
\caption{ (a): 5 second Laue exposure after background subtraction.
The raw data has been binned in 5x5 bins for presentation.
The image plate was 94 cm from the sample and orthogonal to the beam.
The two small boxes indicate the regions shown in the remainder of the figure
while the large box represents the region of Fig. 2b and 2c.
(b) and (c): data around the $(\overline{4} \; \overline{6} \; 0)$ (top) and
$(\overline{3} \; \overline{7} \; \overline{4})$ (bottom) Bragg peaks.
Possible
$<0.1 \; \overline{0.1} \; \overline{0.1}>$
satellite reflections, shown by asterisks,
lie at the vertices of the distorted hexagons.}
\end{figure}

\begin{figure}
\caption{Binned image region with predicted Bragg peak positions marked by
circles.  Careful analysis reveals that all of the data are well accounted for
by the $P6_3/mmc$ calculation.}
\end{figure}

\begin{figure}
\caption{Main panels: Raw data contour plots of the scattering about
the $(\overline{4} \; \overline{6} \; 0)$ Bragg peak.  The isolevels are spaced
by powers of two and the minimum contour level of 32 IU is about twice the
noise in the background.  The region shown is identical to that shown in the
Fig. 2b.  The right panel has a "fake" satellite peak of integrated intensity
500 IU added at
modulation wavevector
$\vec{\delta q}$ = $(0.1 \; \overline{0.1} \; \overline{0.1})$, demonstrating
thata satellite of this intensity would be clearly visible.
Top panels: Slices of the data along the dotted
horizontal lines showing the fluctuations in the background and the relative
magnitude of the "satellite" peak.}
\end{figure}

\begin{figure}
\caption{Binned Laue diffraction data taken in transverse geometry, i.e., with
the image plane at 90 degrees with respect to the incident x-rays.  No signs of
a modulation were found in this data.}
\end{figure}

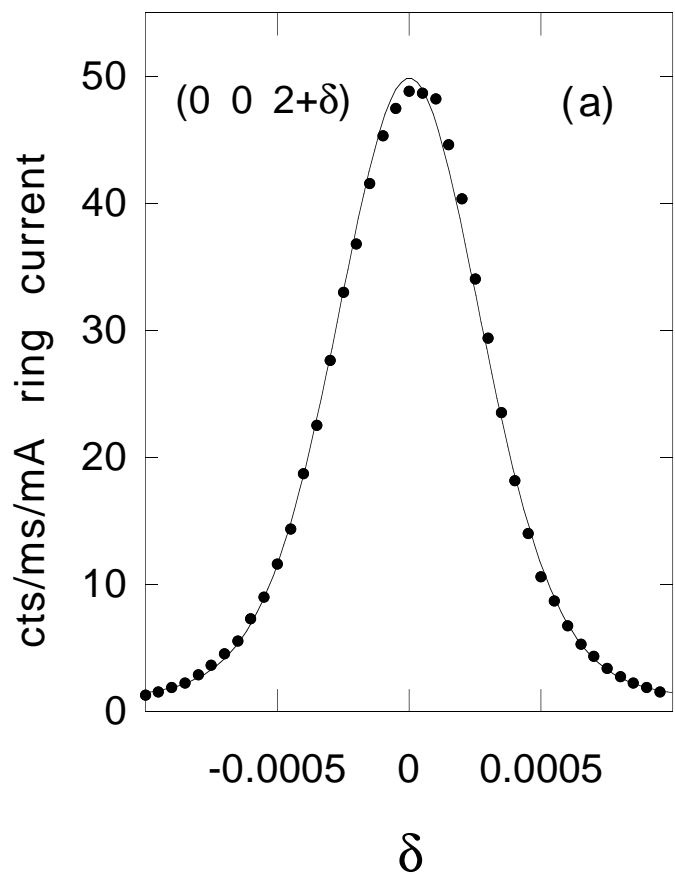
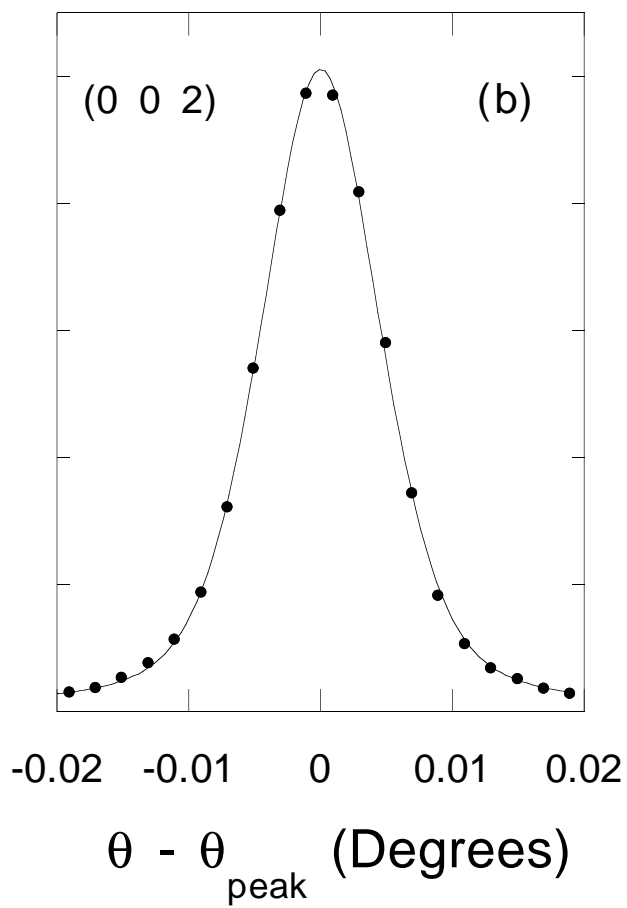
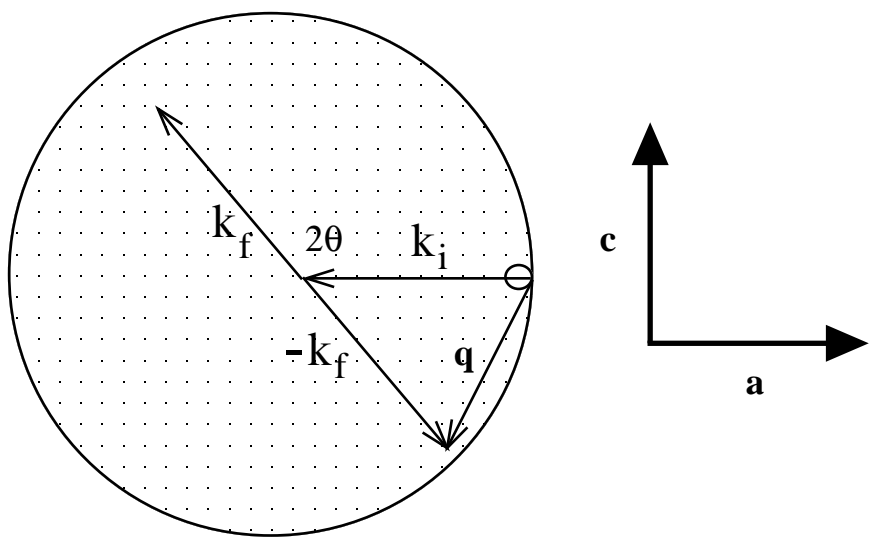
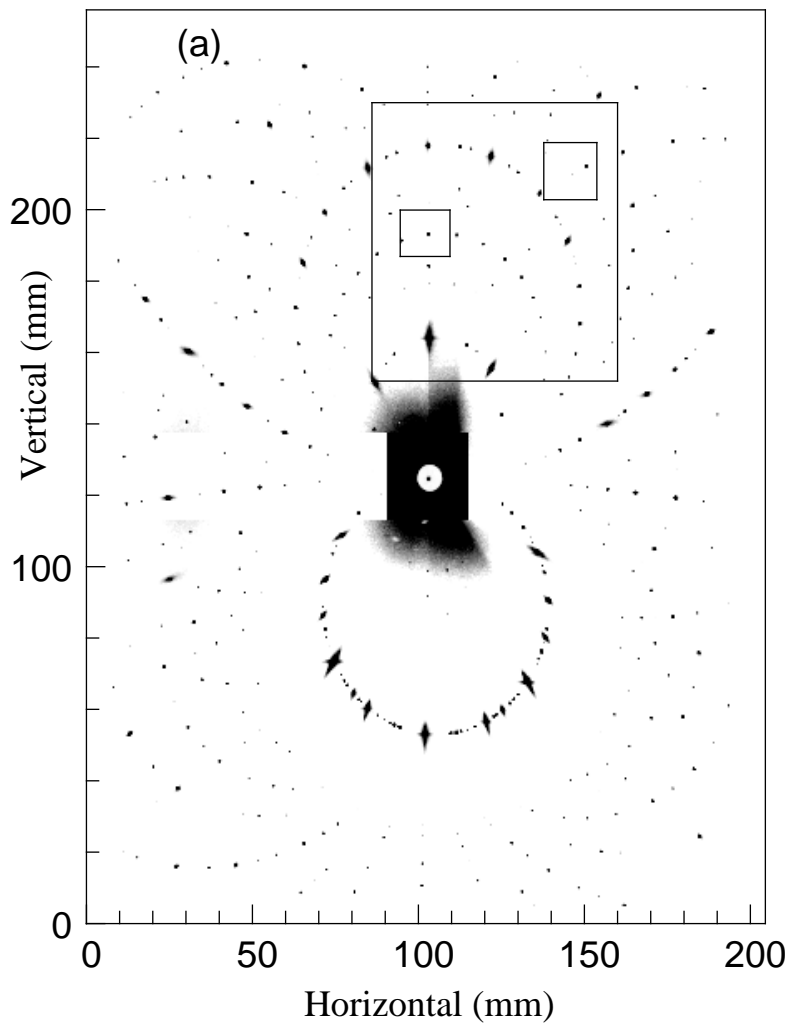
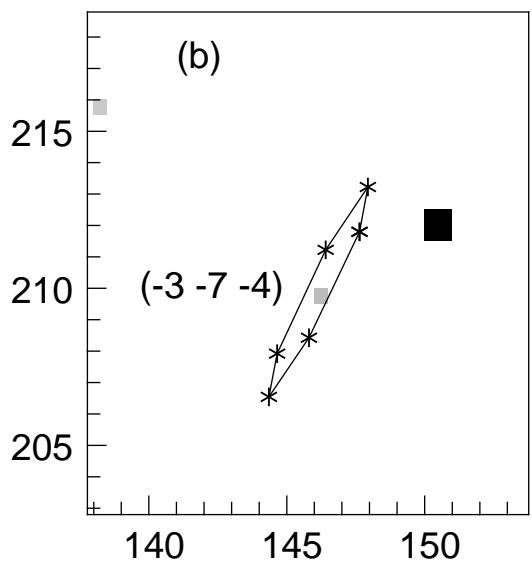
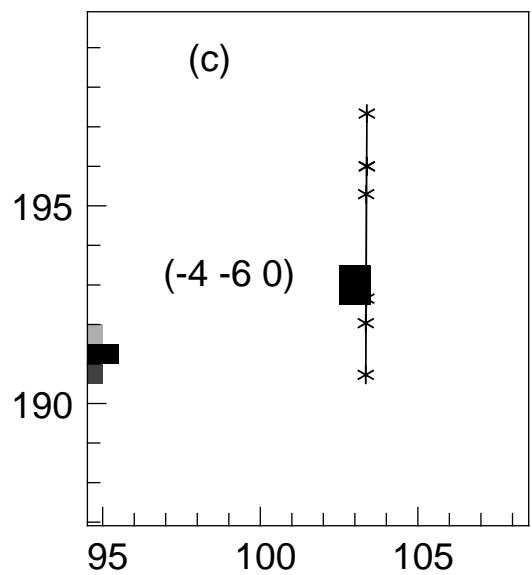
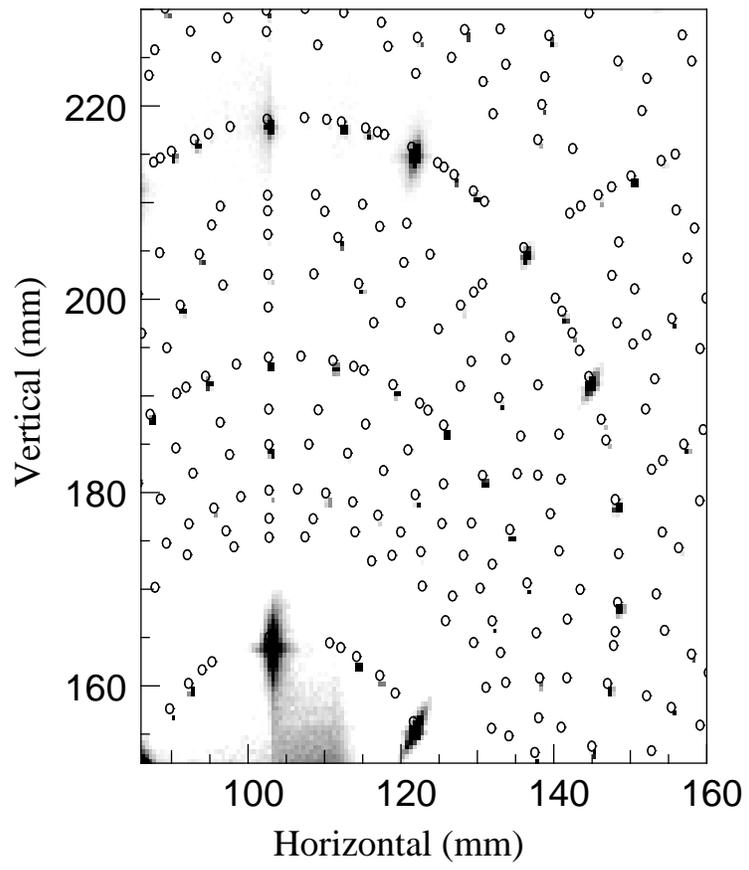
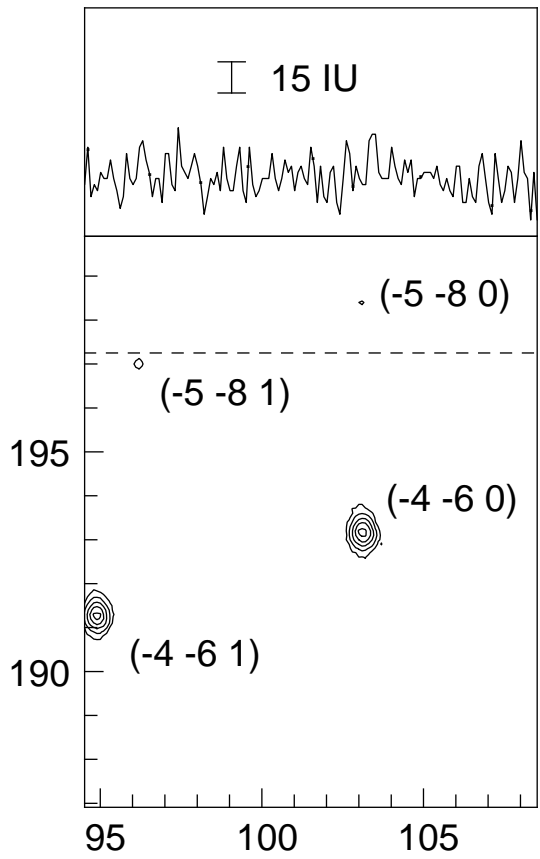
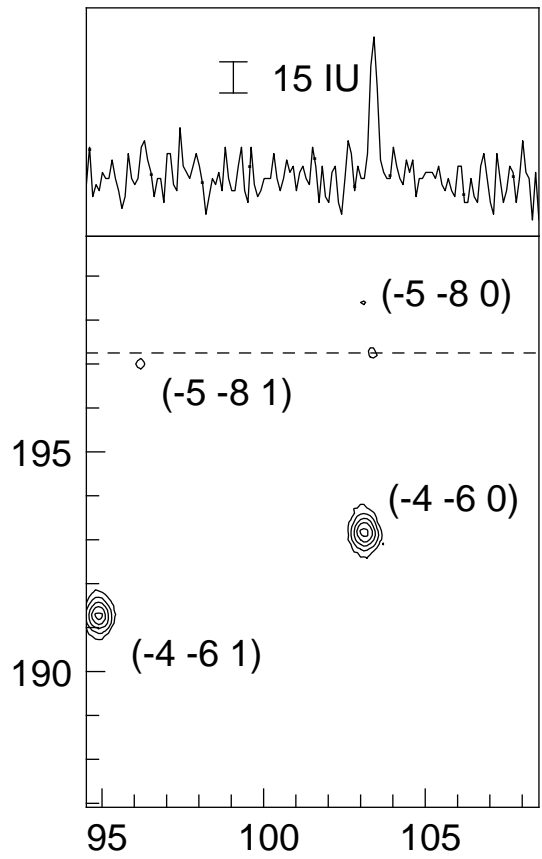
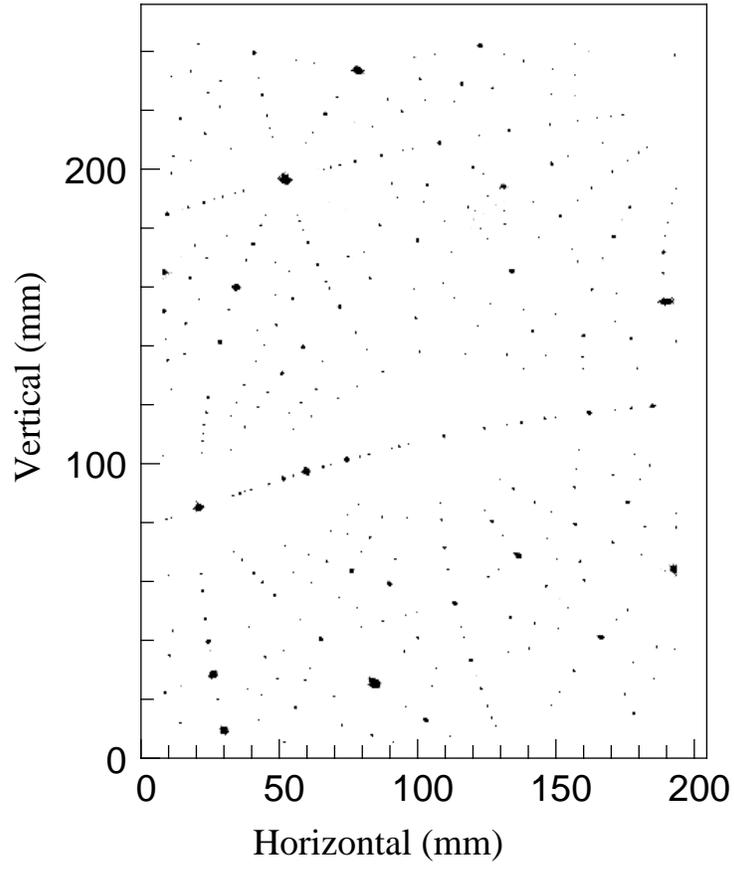
\end{document}